\documentclass[conference]{IEEEtran}
\IEEEoverridecommandlockouts
\usepackage{cite}
\usepackage{amsmath,amssymb,amsfonts}
\usepackage{algorithmic}
\usepackage{graphicx}
\usepackage{textcomp}
\usepackage{xcolor}
\def\BibTeX{{\rm B\kern-.05em{\sc i\kern-.025em b}\kern-.08em
    T\kern-.1667em\lower.7ex\hbox{E}\kern-.125emX}}

\renewcommand{\gcd}{\mathrm{gcd}}

\newcommand{\supp}{\mathrm{supp}}

\newcommand{\X}{{\mathbf X}}
\newcommand{\bL}{\mathbf {\Lambda}}
\newcommand{\N}{{\mathbb N}}
\newcommand{\Z}{{\mathbb Z}}

\newcommand{\F}{{\mathbb F}}

\newcommand{\fF}{\mathfrak F}
\renewcommand{\L}{{\mathbb L}}
\newcommand{\D}{\mathcal{D}}
\newcommand{\R}{\mathcal{R}}

\renewcommand{\H}{\mathcal{H}}
\newcommand{\B}{\mathcal{B}}
\newcommand{\I}{\boldsymbol{\mathcal I}}

\newcommand{\tq}{\;\mid\;}

\newcommand{\bs}[1]{\boldsymbol{#1}}

\newtheorem{theorem}{Theorem}[section]

\newtheorem{proposition}[theorem]{Proposition}
\newtheorem{definition}[theorem]{Definition}
\newtheorem{definitions}[theorem]{Definitions}

\newtheorem{remark}[theorem]{Remark}
\newtheorem{remarks}[theorem]{Remarks}

\newtheorem{example}[theorem]{Example}

\begin{document}

\title{Hyperbolic Sets in Incomplete Tables\\
\thanks{Supported by the Spanish Government under Grant PID2020-113206GB-I00 funded by MCIN/AEI/10.13039/501100011033}
}

\author{\IEEEauthorblockN{Jos\'e Joaqu\'{\i}n Bernal and Juan Jacobo Sim\'on}
\IEEEauthorblockA{\textit{Departamento de Matemáticas} \\
\textit{Universidad de Murcia}\\
Murcia, Spain \\
josejoaquin.bernal@um.es and jsimon@um.es}
}

\maketitle

\begin{abstract}
In this paper, we extend results about the implementation of the Berlekamp-Massey-Sakata algorithm on data tables having a number of unknown values.
\end{abstract}


\section{Introduction.} 

In \cite{BS1} we solved two open problems about the implementation of the Berlekamp-Massey-Sakata algorithm (BMSa, for short) for a class of Abelian Codes called Hyperbolic-like Codes. The first one was to improve the general framework of locator decoding in order to apply it on such abelian codes. The second one was to find sufficient conditions to guarantee that the minimal set of polynomials given by the BMSa is exactly a Groebner basis of the locator ideal.

Now we deal with another classical problem that we call ``the problem of incomplete tables with hyperbolic sets of known values''. Let us introduce it through locator decoding in abelian codes, that we describe briefly.

Let $C$ be a semi simple bivariate abelian code of length $r_1r_2=l\in\N$, over a finite field $\F$. As it is usual, we represent the codewords $c\in C$, as polynomials, $c(X_1,X_2)$. Given a pair $\bs\alpha=(\alpha_1,\alpha_2)$ of primitive $r_1$ and $r_2$ roots of unity, it is known that the code $C$ is totaly determined by its defining set \cite[p. 142]{Huffman-Pless}, $\D_{\boldsymbol{\alpha}}(C) = \left\{ m\in \Z_{r_1}\times\Z_{r_2} \tq f(\boldsymbol{\alpha}^{m})=0, \forall f\in C\right\}$.

Suppose that a codeword $c\in C$ was sent and an element $f=c+e \in \F[X_1,X_2]/(X_1^{r_1}-1,X_2^{r_2}-1)$ has been received. We want to find $e(X_1,X_2)$. To do this, we first consider (theoretically) the array, $U=(u_n)_{n\in \Z_{r_1}\times\Z_{r_2}}$ of syndrome values; $u_n=e(\bs\alpha^n)$. Then we apply the BMSa on $U$ and we get a Groebner basis for the ideal $\bL(U)$ (see Definitions~\ref{conj e ideal de generadores}); from it we find the support (or location positions) of $e$. Finally, by solving a system of linear equations we get the coefficients of $e$.

Clearly, since we know $f$ but not $e$, we only know the syndrome values of $e(X_1,X_2)$ for which $n\in\D_{\bs\alpha}(C)$ because $f(\bs\alpha^n)=e(\bs\alpha^n)$. That is, we only may form an incomplete table of syndrome values. In \cite{BS1}, it is shown that for a Hyperbolic-like code, $C$, if $\D_{\bs\alpha}(C)$ contains an hyperbolic set of amplitude $\delta\leq d(C)$ (see Definition~\ref{el B de delta}), denoted $\B(\delta)$, then the BMSa, and thus locator decoding, may be implemented successfully. An additional problem occurs when one miss some values $u_n\in U$ with $n\in \B(\delta)$. An interesting solution of this problem may be found in the context of Algebraic Geometric Codes in which it is possible to define the Feng-Rao bound (see \cite[Chapter 10]{Cox et al Using}) by the so called Feng-Rao Majority Voting Procedure.

In this paper, we are interested in going beyond the context of locator decoding by considering incomplete data tables of values over an extension field of $\F$ containing $\alpha_1$ and $\alpha_2$; that is, an array of size $r_1\times r_2$, say $\H$. We want to know if $\H$ may be interpreted as a table of syndrome values of a polynomial $e(X_1,X_2)$, as above.



\section{Preliminaries.}

We shall follow the notation used in \cite{BS1}. Throughout this paper $q,\,r_1,\,r_2$ will be positive integers such that $q$ is a power of a prime number, $\gcd(q,r_1r_2)=1$ and $\F=\F_q$ a finite field of $q$-elements. As it is well-known, in this case the polynomial quotient ring $\F(r_1,r_2)= \F[X_1,X_2]/(X_1^{r_1}-1,X_2^{r_2}-1)$ is semi simple. An \textbf{abelian code} is an ideal in $\F(r_1,r_2)$. We set $\I:=\Z_{r_1}\times \Z_{r_2}$ where we only consider canonical representatives. 

We write polynomials as it is usual:  $f \in  \F[X_1,X_2]$ is $f=f(\mathbf X)=\sum a_m \mathbf{X}^m$, where $m=(m_1,m_2)\in \N\times \N$ (we assume $\N=\N\cup \{0\}$) and $\mathbf{X}^m=X_1^{m_1}\cdot X_2^{m_2}$. We write the canonical representatives $f \in  \F(r_1,r_2)$ as $f=\sum a_m \mathbf{X}^m$, where $m=(m_1,m_2)\in \I$. Given $f \in \F[\X]$, we denote by $\overline{f}$ its image under the canonical projection onto $\F(r_1,r_2)$, when necessary. The weight of any  $f\in \F(r_1,r_2)$ will be $\omega(f)=|\supp(f)|$. 

For each $i\in \{ 1,2\}$, we denote by $R_{r_i}$ (resp. $\R_{r_i}$) the set of all $r_i$-th roots of unity (resp. primitive $r_i$-th roots) and we define $R=R_{r_1}\times R_{r_2}$ ($\R=\R_{r_1}\times \R_{r_2}$). We denote by $\L|\F$ an extension field that contains $R_{r_i}$, for $i=1,2$.

For any $f=f(\X) \in \F[\X]$ and $\boldsymbol{\alpha}=(\alpha_1,\alpha_2)\in R$, we write $f(\boldsymbol{\alpha})=f(\alpha_1,\alpha_2)$ and for any $m=(m_1,m_2)\in \I$, we write  $\boldsymbol{\alpha}^{m} = (\alpha_1^{m_1},\alpha_2^{m_2})$.

 
 In what follow we shall review some results in \cite{Sakata} and \cite{BS1} that will be needed.
 
 We consider the partial ordering in $\Sigma_0$ given by $(n_1,n_2) \preceq (m_1,m_2) \Longleftrightarrow n_1\leq m_1$ and $n_2\leq m_2.$  On the other hand, we will use a (total) monomial ordering \cite[Definition 2.2.1]{Cox}, denoted by ``$\leq_T$'', as in \cite[Section 2]{Sakata}. This ordering will be either the lexicographic order (with $X_1>X_2$) \cite[Definition 2.2.3]{Cox} or the (reverse) graded order (with $X_2>X_1$) \cite[Definition 2.2.6]{Cox}. The meaning of ``$\leq_T$'' will be specified as required. 
 
 Given $s\in\N\times\N$ we define $\Sigma_s=\{n\in \N\times\N\tq s\preceq n\}$. In particular, $\Sigma_0=\N\times\N$.  We recall that $\N=\N\cup \{0\}$.
  
 \begin{definition}\label{def tabla de sindromes}
   Let $e\in\F(r_1,r_2)$, $\tau\in I$ and $U=(u_n)_{n\in \Sigma_0}$ an array such that $u_n=e(\boldsymbol{\alpha}^{\tau+n})$. We shall call \textbf{syndrome values} to those $e(\boldsymbol{\alpha}^{\tau+n})$ and  $U$ will be called the \textbf{syndrome table afforded by $\tau$ and $e$}.
 \end{definition}

Note that any syndrome table as defined above verifies that if  $n=(n_1,n_2),\;m=(m_1,m_2) \in \Sigma_0$ are such that $n_i\equiv m_i\mod r_i$ for $i=1,2$ then $u_n=u_m$; that is, $U$ is are what it is called \textbf{doubly periodic arrays} of period $r_1\times r_2$ (see \cite[p. 324]{Sakata}) and, clearly, they are totally determined by its indexes over the range $0\leq n_1\leq r_1-1$ and $0\leq n_2\leq r_2-1$. In practice, we consider finite tables.

The BMSa is an iterative procedure over the array $U$ with respect to a monomial ordering over $\I$; although theoretically, as we will see later, we consider evaluations over all $\Sigma_0$.

In order to make iterations over $\I$ (viewing its elements in $\Sigma_0$) we have to specify the notion of successor (see \cite[p. 137]{BS1} and \cite[p. 322]{Sakata}). In the case of graded order we have the usual one
$$l+1=\begin{cases}
	    (l_1-1,l_2+1) & \text{if } l_1>0\\
            (l_2+1,0) & \text{if } l_1=0
      \end{cases}.$$
In the case of lexicographic order we introduce a modification
$$l+1=\begin{cases}
	    (l_1,l_2+1) & \text{if } l_2<r_2-1\\
            (l_1+1,0) & \text{if } l_2=r_2-1
      \end{cases}.$$

As the reader will see later, by Theorem~4 in \cite{BS1}, we only have to make iterations into sets of indexes of the form  $\B(2t+1)$ (see Definition \ref{el B de delta}), for certain $t\in\N$. Now we relate polynomials and linear recurring relations, and then we contruct some associated sets and ideals. We follow \cite{BS1} (see also \cite[Section 2]{Sakata}). As in \cite[p. 323]{Sakata} for any $f\in \F[\X]$, we denote by $LP(f)$ the \textbf{leading power product exponent} (or multidegree) of $f$ with respect to $<_T$.

\begin{definition}
	Let $U$ be a doubly periodic array, $f\in \L[\mathbf X]$, with multidegree $LP(f)=s$ and $n\in \Sigma_0$. We write $f=\sum_{m\in \supp(f)}f_m\mathbf{X}^m$ and define 
	\[f[U]_n=\begin{cases}
		\displaystyle{ \sum_{m\in\supp(f)}f_m u_{m+n-s}}& \text{if }  n\in\Sigma_s\\
		0 & \text{otherwise}
	\end{cases}.\]
	
	The equality $f[U]_n=0$ will be called a \textbf{linear recurring relation}.
\end{definition}

\begin{definition}
 For $l\in \Sigma_0$, we define the finite subarray $u^l=\{u_m\tq m <_T l\text{ y } m\prec (r_1,r_2)\}$. 

Note that $u^l$ is a finite set.
\end{definition}

  \begin{definition}\label{conj e ideal de generadores}
Let $u^l\subseteq U$ be a finite subarray and  $f\in \L[\mathbf X]$ with $LP(f)=s$.
 \begin{enumerate}
 \item We say that $f$ generates  $u^l$ if $f[U]_k=0$ in each pair $k$ such that $s\preceq k$ and $k<_T l$, and we write $f[u^l]=0$. If $\{k\in \Sigma_0\tq  s\preceq k\;\text{ y }\;k<_T l\}=\emptyset$ we will also write $f[u^l]=0$.
 
  \item We denote the \textbf{set} of generating polynomials for  $u^l$ as 
  \[\bL(u^l)=\{f\in \L[\mathbf{X}]\tq f[u^l]=0\}.\]  
  
  \item We say that $f$ generates $U$ if $f[U]_l=0$ on each pair $l\in \Sigma_0$ and we write $f[U]=0$.   
  
\item We denote the \textbf{ideal} of generating polynomials for $U$ as 
  $$\bL(U)=\left\{f\in\L[\mathbf{X}] \tq f[U]=0\right\}.$$
\end{enumerate}
  \end{definition}

\section{The Berlekamp-Massey-Sakata algorithm.}

The BMSa is an iterative procedure over $\I$ in which at each step, $l$, two sets of polynomials, called $G_l$ and $F_l$, are updated until $F_l$ gets a Groebner basis for $\bL(U)$ and hence a system of linear recurring relations.

Let us give a brief description of the algorithm. 


Let $U$ be a doubly periodic array, $u^l\subseteq U$ and consider a set of polynomials $F_l=\left\{f^{(1)},\dots,f^{(d)}\right\}\subset\bL(u^l)$. Set $s^{(i)}=LP(f^{(i)})$, where $s^{(i)}=(s^{(i)}_1,s^{(i)}_2)$.

\begin{definitions} Taking the notation as above.
 \begin{enumerate}
 \item The footprint (or the connection footprint \cite[p. 1621]{Blah}) of $\bL(u^l)$ is the set 
 \begin{eqnarray*}
\Delta(u^l)&=&\{n\in \Sigma_0\tq n\preceq(s^{(i)}_1-1,s^{(i+1)}_2-1)\\
&&\text{for some } 0\leq i\leq d-1\}.
 \end{eqnarray*}
  \item We say that $F_l$ is a minimal set of polynomials for $u^l$ if
  \begin{enumerate}
   \item The sequence $s^{(1)},\dots,s^{(d)}$  (called defining points) satisfies  
   \[ \hspace{-.7cm}s^{(1)}_1>\ldots>s^{(d)}_1=0 \quad\text{and}\\
    \quad 0=s^{(1)}_2<\ldots<s^{(d)}_2.\]
    \item If $g\in\L[\X]$ is such that $LP(g)\in \Delta(u^l)$ then $g\not\in\bL(u^l)$ (that is, $g[u^l]\neq 0$).
  \end{enumerate}
 \end{enumerate}
\end{definitions}

If $F$ is a Groebner basis of $\bL(U)$ we write $\Delta(U)$ to denote its corresponding footprint.

In \cite[Section 4]{Sakata}, it is proved that every set $\bL(u^l)$ has at least one minimal set of polynomials, $F_l$. Moreover, if $l,l'\in \I$ are such that $l<_T l'$ then $\bL(u^{l'})\subseteq \bL(u^{l})$ and the footprints verify $\Delta(u^l)\subseteq\Delta(u^{l'})$. 

As we said before there exists another set of polynomials that we call 
\begin{equation}\label{Conjunto de los G}
  G_l=\{g^{(i)}\tq i=1,\dots,d-1\}
\end{equation}
They are used in each iteration to update minimal sets of polynomials and, in turn, they will be updated as well. More specifically, there are pairs $k_1,\dots,k_{d-1}$ in $\I$ such that, for $i=1,\dots,d-1$, $k_i<_T l$, $g^{(i)}[u^{k_i}]_{k_i}\neq 0$ and $k_i-LP(g^{(i)})=(s^{(i)}_1-1,s^{(i+1)}_2-1)$ (see \cite[p. 327]{Sakata}). 
	

Now, let us see a brief sketch of the algorithm. Following \cite[p. 331]{Sakata} (see also \cite[Paragraph 4.1]{BS1}), we start with a doubly periodic array of period $r_1\times r_2$, say $U=(u_n)_{n\in \Sigma_0}$, where we have already defined a partial order $\preceq$ and a monomial ordering $\leq_T$, together with a notion of successor for $\I$.

\begin{itemize}
    \item We initialize $F_{(0,0)}=\{1\}$ and $G_{(0,0)}=\emptyset=\Delta(u^{(0,0)})$.
  \item For $l=(0,0)$, we have already the initializing objects.  
  \item Now, for $l\in \I$ with given $F_{l}$, $G_l$ and $\Delta(u^{l})$ (including $l=(0,0)$), we update them as follows:    
  \item For each $f\in F_l$ we compute $f[u^{l+1}]_l$. Then  
  \begin{enumerate}
   \item If $f[u^{l+1}]_l=0$ for all $f\in F_l$ then $F_l=F_{l+1}$,  $G_l=G_{l+1}$ and $\Delta(u^{l})=\Delta(u^{l+1})$; so that there is no strictly updating.  
   \item Otherwise, $F_l\neq F_{l+1}$.
   \begin{itemize}
    \item In this case, each $f\in F_l$ such that $f[u^{l+1}]_l\neq 0$ will be replaced following what is called \textbf{the Berlekamp procedure} \cite[Lemma 6]{Sakata} or \cite[Theorem 1]{BS1} (so that $F_l\neq F_{l+1}$). To make these replacement we use the corresponding auxiliar set $G_l$.
   \end{itemize}

  \end{enumerate}  
Finally, at the end of each step, in case $F_l\neq F_{l+1}$, some of the replaced polynomials of $F_l$ will be used to update $G_l$ to $G_{l+1}$ \cite[Theorem 2]{Sakata}.

With respect to the footprint, it may happen:
    \begin{enumerate}
  \item If $l-LP(f)\in\Delta(u^l)$ for all $f\in \F_l$, then $\Delta(u^{l})=\Delta(u^{l+1})$ and  $G_l=G_{l+1}$ \cite[Theorem 1]{Sakata}.
  
  \item If $l-LP(f)\not\in\Delta(u^l)$ for some $f\in \F_l$, then $\Delta(u^{l})\varsubsetneq \Delta(u^{l+1})$; in fact $l-LP(f)\in\Delta(u^{l+1})$. 
 \end{enumerate}
 \end{itemize}

\subsection{Termination criteria}

Let $U$ be a syndrome table afforded by  $\tau$ and $e$, with $\omega(e)\leq t$. In \cite{Sakata} Sakata and other authors of modern versions (see \cite{Cox et al Using,Sakata 3}) prove that if a enough number of steps are covered, the updates of minimal sets of polynomials will finish by getting a Groebner basis for  $\bL(U)$; that is, the last minimal set of polynomials obtained is a Groebner basis and the last footprint obtained is its footprint in the classical sense, as in \cite{Cox} (see also \cite[p. 1615]{Blah}

The problem of finding a minimal number of steps in the implementation of the BMSa was solved in \cite{BS1} for a class of hyperbolic like codes (see also \cite{BS2} for the general case). From Theorem 3 and Theorem 4 in \cite{BS1} we have the result showed in Theorem \ref{implementacion del aBMS en el conjunto hiperbolico}. 

The following result is essential in all the results.

\begin{proposition}\label{propiedad-cardinal-delta}
Let $e\in\F(r_1,r_2)$. If $U$ is the syndrome table afforded by $e$ and some $\tau\in\I$ then $|\bL(U)|\leq t$.
\end{proposition}

We need the following definition.
 \begin{definition}{(See \cite[p. 136]{BS1}).}\label{el B de delta}
Let $r_1$, $r_2$ as above and $\delta\in\N$. We shall call hyperbolic set of amplitude $\delta$ to a subset of pairs in $\I$ such that 
\begin{eqnarray*}
\B(\delta)&=&\left\{(l_1,l_2)\in \I\tq (l_1+1)(l_2+1)\leq \delta\right\}\setminus\\ 
&& \left\{(\delta-1,0),(0,\delta-1)\right\}
\end{eqnarray*}
 \end{definition}
 
\begin{theorem}\label{implementacion del aBMS en el conjunto hiperbolico}
 Let $U$ be a syndrome table afforded by $\tau$ and $e$, with $\omega(e)\leq t\leq 4$, such that $u_{(0,j)}\neq 0$, for some $j<t$ (respectively, $u_{(i,j)}\neq 0$ for some $i+j=t$). Then, in order to obtain a Groebner basis for $\bL(U)$, using the lexicographical ordering (respectively, the graduate ordering), it is enough to implement the BMSa on the set of indexes $l\in\B(2t+1)$.
\end{theorem}

\begin{remark}
It can be proved that the exclusion of even the ``last point'' (with respect to the corresponding ordering) in $\B(2t+1)$ may produce the failure in the implementation.
\end{remark}


\subsection{The locator ideal.}\label{paragrafo locator ideal}
 
Let $U$ be a syndrome table afforded by $\tau$ and $e$. As the reader may suppose, the implementation of the BMSa only make sense when the polynomial $e \in \F(r_1,r_2)$ is not known, as in the context of locator decoding. So, once one has implemented the BMSa obtaining the system of linear recurring relations that generates it, and hence the Groebner basis for the ideal $\bL(U)$ of $\L[X_1,X_2]$, the next task is to find $e(X_1,X_2)$. Let us comment some theoretical results (in the context of the ring $\L(r_1,r_2)$) and the procedure to find, first, the support of the error, $\supp(e)$, and finally their coefficients.
 
As we have commented in the Introduction (in the context of abelian codes) given a fixed $\boldsymbol{\alpha}\in \R$, any ideal  $I$ in $\L(r_1,r_2)$ is totally determined by its defining set, $\D_{\boldsymbol{\alpha}}(I) = \left\{ m\in \I \tq f(\boldsymbol{\alpha}^{m})=0, \forall f\in I\right\}$. It is also well-known that defining sets may be considered for sets of polynomials and, moreover, if $G\subset \L(r_1,r_2)$ then
 $\D_{\boldsymbol{\alpha}}(G)=\D_{\boldsymbol{\alpha}}(\langle G\rangle)$.
 
 There is an important relation between defining sets and footprints (in the classical sense for ideals \cite{Cox}) that we may find in \cite[Remark 2]{BS1} (together with other interesting properties); to wit
 $ |\Delta(I)|=|\D_{\boldsymbol{\alpha}}(I)|$.
 
 \begin{definition}{\cite[Definition 6]{BS1}}
 Let $e$ be a polynomial defined in a subfield of $\L$. The locator ideal is
  $L(e)=\left\{f\in \L(r_1,r_2)\tq f(\boldsymbol{\alpha}^n)=0,\;\forall n\in \supp(e)\right\}$.
 \end{definition}

Clearly, as $\alpha_1,\alpha_2\in \L$, we have that $\D_{\boldsymbol{\alpha}}(L(e))=\supp(e)$; the support of $e$. 
 
Let $U$ be a syndrome table afforded by $\tau$ and $e$. We denote by $\overline{\bL(U)}$ the canonical projection of $\bL(U)\leq \L[\X]$ onto the ring $\L(r_1,r_2)$. The next result explains us the relationship between the ideals $\bL(U)$ and $L(e)$.

\begin{theorem}{\cite[Theorem 2]{BS1}}\label{el locator y el delta}
In the setting above, we have $\overline{\bL(U)}=L(e)$.

Hence $\D_{\boldsymbol{\alpha}}(\overline{\bL(U)})=\D_{\boldsymbol{\alpha}}(L(e))=\supp(e)$.
\end{theorem}

Finally, under the assumption $\omega(e)\leq t$, the size of $\B(2t+1)$ clearly has enough elements to solve the suitable linear system of equations to find all coefficients of $e$.

\section{The BMSa on hyperbolic tables.}\label{construccion de las relaciones y el polinomio}
%

Let $\F$, $\L$, $\I$, $t,r_1,r_2$ and $e$ be as in the previous sections, with  $\omega(e)\leq t$, for $i=1,2$.


\begin{definition}\label{def tablas hiperbolicas}
 Let $\H=(h_n)_{n\in \I}$ be an incomplete data table with values in $\L$. We say that $\H$ is an \textbf{hyperbolic table} if there exist an hyperbolic set of amplitude $\delta=2t+1$, for some element $\tau\in \I$ such that $h_n$ is a known value for all $n\in \tau+\B(2t+1)$.
\end{definition}

Observe that the condition $\tau+\B(2t+1)\subseteq \I$ implies that necessarily $t\leq \lfloor\frac{r_i}{2}\rfloor$, for $i=1,2$. So, from now we assume that $\omega(e)\leq t\leq \lfloor\frac{r_i}{2}\rfloor$, for $i=1,2$. 
.

For a given incomplete table $\H$,  our first task is to determine if there exists an hyperbolic set of amplitude $\delta=2t+1$, for some $t\in \N$ and an element $\tau\in \I$, in such a way that $\H$ is a hyperbolic table. 

\begin{example}
Let $\F=\F_2$ and $\L=\F_{2^4}$, $r_1=r_2=5$. Set
\begin{equation}\label{ejemplo matriz incompleta}
 \H=\begin{pmatrix}
*&a^5 & a^{10} & a^{10} & a^{5}\\ 
* & a^{4} &a^4& *&0\\
a^{13} & a^{13} & * &*&a^8\\
a^7 & a^2&0&a^2&a^7\\
a^{11}& 0 &* & *&a
\end{pmatrix}.
\end{equation}

Recall that we do not know if $\H$ is a table of syndrome values afforded by a polynomial $e\in \F(r_1,r_2)$.


The next step will be to consider an array $U$ of size $5\times 5$ that agrees with $\H$ in the positions in $\B(2t+1)$. 
\[U=\begin{pmatrix}
\boldsymbol{a^5} & \boldsymbol{a^{10}} & \boldsymbol{a^{10}} & \boldsymbol{a^{5}}&*\\ 
\boldsymbol{a^{4}} &\boldsymbol{a^4}& *&0 &*\\
 \boldsymbol{a^{13}} & * &*&a^8&*\\
 \boldsymbol{a^2}&0&a^2&a^7&*\\
 0 &* & *&a&*
\end{pmatrix}.\]

To be consistent we use the known values of $\H$ in $\I\setminus\B(2t+1)$, but they will not be used. We are trying to apply the BMSa on the array $U$.

\end{example}

\bigskip

So, let us begin with an incomplete hyperbolic table with size $r_1\times r_2$, and let $t,\tau\in \I$ such that $\tau+\B(2t+1)$ are indexes that corresponds to known values. Then, we define the array $U=(u_n)$ such that $u_{(i,j)}=h_{\tau+(i,j)}$ for any $(i,j)\in \B(2t+1)$.

In order to apply Theorem \ref{implementacion del aBMS en el conjunto hiperbolico} we know that if $U$ is a syndrome table afforded by some $\tau$ and $e$, with $\omega(e)\leq t\leq 4$ then, by applying the BMSa on $U$ we find $F$ a Groebner basis for the ideal $\bL(U)$. Let us denote $\Delta=\Delta(F)=\Delta(U)$.

Therefore, if $\H$ appears to be a syndrome table afforded by a polynomial $e\in \F(r_1,r_2)$ and $\tau'$, with $\omega(e)\leq t\leq 4$, then $U$ appears to be a syndrome table afforded by $e$ and $\tau+\tau'$. So, we can find a Groebner basis for $\bL(U)$, $F$. From it we obtain $D_\alpha(\langle F\rangle)=\supp (e)$ and we find the coefficients of $e$ because we have enough known values (equations) in $\B(2t+1)$. Once we have found $e$ we can recover any unknown value in $U$ and $\H$ (observe that since $e\in\F(r_1,r_2)$ the arrays are double periodic).

\begin{remarks}
Through the application of the BMSa we have two different criteria to determine that we will not be succeed, to wit:
\begin{enumerate}
	\item If, in any step, we obtain $\Delta(u^l)$ with cardinality greater than $t$ we can conclude that our array $U$ can not be afforded by any polynomial $e$ with $\omega(e)\leq t$ (see Proposition \ref{propiedad-cardinal-delta}). Since, we can assume that we are taking the greatest $t$ such that $\H$ is an hyperbolic array, we stop the algorithm. 
	\item It can be proved that $X^{r_i}-1$ belongs to $\bL(U)$ for $i=1,2$. So, when we have finished the algorithm on the values in $\B(2t+1)$ we obtain a set of polynomials $F$. Since we do not know that $\H$ is an array afforded by a polynomial with weight less than or equal to $t$, we can not assure that $F$ is a Groebner basis for $\bL(U)$. Moreover, if $F$ is not a Groebner basis for $\langle F\cup \{X_1^{r_1}-1,X_2^{r_2}-1\}\rangle$, it surely can not be a Groebner basis for $\bL(U)$. In that case, we may claim that $\H$ is not an array afforded by a polynomial with weight less that or equal to $t$.
	
	\item If the previous cases do not occur, we obtain a candidate $e$ and we check if the array $\H$ is in fact afforded by it. In case that $\H$ is afforded by a polynomial with weight less than or equal to $t$ it must be $e$.
\end{enumerate}
\end{remarks}

\section{Estimation of unknown values with indexes inside hyperbolic sets.}

Suppose that, for an incomplete table $\H=(h_n)_{n\in \I}$ we construct a new array $U$ under some parameters $\tau$ and $t$, to implement the BMSa as it has done with the table in Equation~\eqref{ejemplo matriz incompleta}; however there are some (few) pairs $n\in \tau+\B(2t+1)$ for which $h_n$ are unknown values. We want to look for alternatives to find a Groebner basis for $\bL(U)$. As one may expect, we cannot estimate any value of $\H$; in fact we have to restrict ourselves to estimate values having ``border indexes''; that is, for $2\leq t\leq 4$, we consider the set
\[\fF=\left\{(l_1,l_2)\in \B(2t+1)\tq 2t\leq (l_1+1)(l_2+1)\right\}. \]

\begin{theorem}\label{inferencia principal}
Let $\F$, $\L$, $t,r_1,r_2,\tau$ and $e$ be as above, with  $\omega(e)\leq t\leq 4$, for $i=1,2$. Let $U$ be the syndrome table afforded by $e$ and $\tau$. Suppose that, following the BMSa under any of the monomial orders considered, we have constructed, for $l=(l_1,l_2)\in \B(2t+1)$, the sets $F_l$, $\Delta(u^l)$ and $G_l$. 

Suppose that we do not know the value of $u_{l}$.
 
 If $l\in\fF$ is such that $l_1,l_2\neq 0$ then there exists $f\in F_l$ such that $LP(f)\preceq l$ and $f[u^l]_l=0$; except for the following cases:
 
 \begin{enumerate}
  
  \item  $d=2$, $s_1^{(1)}=t$, $s_2^{(2)}=1=l_2$ and the implementation is doing under the inverse graduate ordering ($X_2>X_1$).
  
  \item $d=2$, $s_1^{(1)}=1=l_1$, $s_2^{(2)}=t$ and the implementation is doing under the lexicographic ordering ($X_1>X_2$).
  
  \item $d=2$, $s^{(1)}=(2,0)$, $s^{(2)}=(0,2)$ and $l=(1,3)$. 
  
  \item $d=2$, $s^{(1)}=(2,0)$, $s^{(2)}=(0,2)$ and $l=(3,1)$. 
  
   \item $d=3$, $s^{(1)}=(2,0)$, $s^{(3)}=(0,t-1)$ and $l=(1,t-1)$. 
   
    \item $d=3$, $s^{(1)}=(t-1,0)$, $s^{(3)}=(0,2)$ and $l=(t-1,1)$.
 \end{enumerate}
\end{theorem}

As the next result shows us, in the cases \textit{1) to 6)} considered in the theorem above, the construction of a Groebner basis is possible and so the estimation of the unknown value.

\begin{proposition}\label{salvando los muebles para los casos del Teorema}
 Take all the setting of Theorem~\ref{inferencia principal} above; so that $l=(l_1,l_2)\in \fF$ verifies $l_1,l_2\neq 0$. Then, in the cases considered above there are one or two elements $b,\,c\in \L$ and polynomials $h_b,\,h_c$ satisfying the following properties:
 
 \begin{itemize}
  \item[a)] $h_b\in\bL(u^l), h_b[U]_k=0$ and (if it exists) $h_c\in\bL(u^l), h_c[U]_k=0$ for all $k\in \B(2t+1)$ such that $k\geq_T l$ (hence the only one or the two polynomials will be elements of the Groebner basis obtained by the BMSa)
  \item[b)] For every $k\in \B(2t+1)$ such that $k\geq_T l$, we have the equality $G_l=G_{k}$ 
 \end{itemize}
 
Hence, any subsequent updating of $F_k$, for $k\in \B(2t+1)$, $k\geq_T l$, does not depend on the elements $b,\,c\in \L$ that must be taken to construct $h_b$ and $h_c$ to get a Groebner basis for $\bL(U)$.
\end{proposition}

 Now let us describe the cases and the construction of the polynomials $h_b, h_c$ mentioned in the previous result. For any $g^{(i)}\in G_l$ we set $g^{(i)}[u^{k_i}]_{k_i}=v_i\neq 0$ (see  Remark~\ref{Conjunto de los G}).
 
 \begin{enumerate}
 
 \item Under the inverse graduate ordering ($X_2>X_1$), having $d=2$, $s_1^{(1)}=t$ and $s_2^{(2)}=1=l_2$. The polynomial is $$h_b=f^{(2)}-\frac{b}{v_1}g^{(1)}.$$ 
 
  \item Under the lexicographic ordering ($X_1>X_2$), having $d=2$, $s_1^{(1)}=1=l_1$ and $s_2^{(2)}=t$. The polynomial is $$h_b=f^{(1)}-\frac{b}{v_1}g^{(1)}$$
  
  \item Under any of the orders considered, having $d=2$, $s^{(1)}=(2,0)$, $s^{(2)}=(0,2)$ and $l=(1,3)$.  The polynomial is $$h_b=f^{(2)}-\frac{b}{v_1}g^{(1)}$$
  
  \item Under any of the orders considered, having $d=2$, $s^{(1)}=(2,0)$, $s^{(2)}=(0,2)$ and $l=(3,1)$.  The polynomial is $$h_b=f^{(1)}-\frac{b}{v_1}g^{(1)}$$
  
\item Under any of the orders considered, having $d=3$, $s^{(1)}=(2,0)$, $s^{(3)}=(0,t-1)$ and $l=(1,t-1)$.  The polynomials are
  $$h_b=f^{(2)}-\frac{b}{v_2}g^{(2)} \text{ and } h_c=f^{(3)}-\frac{c}{v_1}g^{(1)}.$$

\item Under any of the orders considered, having $d=3$, $s^{(1)}=(t-1,0)$, $s^{(3)}=(0,2)$ and  $l=(t-1,1)$.  The polynomials are
   $$h_b=f^{(1)}-\frac{b}{v_2}g^{(2)} \text{ and } h_c=f^{(2)}-\frac{c}{v_1}g^{(1)}.$$
 \end{enumerate}

Note that Proposition \ref{salvando los muebles para los casos del Teorema} says that there exist the value $b$ (or values $b$ and $c$)  and the corresponding polynomial $h_b$ (or polynomials $h_b, h_c$) but, in fact, it follows from the proof that it is an unknown value because it depends on the unknown value $u^l$. Later, we will explain how to proceed in order to get the derired Groebner basis. 

Now we finish the estimation of values by considering last points of the axes of  $\B(2t+1)$ which are not considered in Proposition \ref{salvando los muebles para los casos del Teorema}, that is, $l=(2t-1,0)$ or $l=(0,2t-1)$. 

\begin{theorem}\label{inferencia en los extremos de los ejes}
Let $\F$, $\L$, $t,r_1,r_2,\tau$ and $e$ be as above, with  $\omega(e)\leq t\leq 4$, for $i=1,2$. Let $U$ be the syndrome table afforded by $e$ and $\tau$. Suppose that, following the BMSa under any of the monomial orders considered, we have constructed, for $l=(l_1,l_2)\in \B(2t+1)$, the sets $F_l$, $\Delta(u^l)$ and $G_l$. 

Suppose that we do not know the value of $u_{l}$.
 
 If $l\in\fF$ is such that  $l=(2t-1,0)$ or $l=(0,2t-1)$ then there exists $f\in F_l$ such that $LP(f)\preceq l$ and $f[u^l]_l=0$; except for the following cases:
 
 \begin{enumerate}
  \item $l_1=0$, $l_2=2t-1$ and $s_2^{(2)} = t$ 
  \item $l_2=0$, $l_1=2t-1$ and $s_1^{(1)}=t$.
 \end{enumerate}
In both cases, it is $d=2$.
\end{theorem}

\begin{proposition}
Take all setting in Theorem~\ref{inferencia en los extremos de los ejes}. In this case, $G_l$ is a single set $G_l=\{g\}$ associated with $k\in \I$ having $g[u^{k}]_{k}=v\neq 0$ (See  Remark~\ref{Conjunto de los G}).

Then, in the cases considered above there is an element $b\in \L$ and a polynomial $h_b$ satisfying the following properties:
 
 \begin{itemize}
  \item[a)] $h_b\in\bL(u^l), h_b[U]_k=0$ for all $k\in \B(2t+1)$ such that $k\geq_T l$ (hence the $h_b$ will be an element of the Groebner basis obtained by the BMSa).
  \item[b)] For every $k\in \B(2t+1)$ such that $k\geq_T l$, we have the equality $G_l=G_{k}$.
 \end{itemize}
 
Hence, any updating of $F_k$, for $k\in \B(2t+1)$ such that $k\geq_T l$, does not depend of the $b\in \L$ chosen to construct $h_b$ to get a Groebner basis for $\bL(U)$.

 Let us describe the cases and the construction of the polynomial through the implementation of the BMSa, under any of the orders considered:

\begin{enumerate}
 \item Having $l_1=0$, $l_2=2t-1$ and $s_2^{(2)}=t$.   The polynomial is
 $$h_b=f^{(2)}-\frac{b}{v}g^{(1)}.$$ 
 
 \item Having $l_2=0$, $l_1=2t-1$ and $s_1^{(1)}=t$. The polynomial is
 $$h_b=f^{(1)}-\frac{b}{v}g^{(1)}.$$
\end{enumerate}
\end{proposition}

To finish let us explain how one may proceed to apply the results above to find the desired Groebner basis for $\bL(U)$.\\

Take all notation above and suppose we are implementing the BMSa until we reach step $l\in \fF$ for which we do not know the value $u_l$ of $U$.
\begin{itemize}
 \item[i)] If $l$ does not corresponds to any cases considered in Theorem~\ref{inferencia principal} \textit{1) to 6)} or Theorem~\ref{inferencia en los extremos de los ejes} \textit{1) to 2)} then as there is $f\in F_l$ for which $LP(f)\preceq l$ we may know the value $u_l$ by considering the equation $f[u^l]_l=0$.
 \item[ii)] Otherwise, having in mind that the polynomials $h_b$ or $h_{(b,c)}$ do not intervene in subsequent steps, we may continue the implementation of the BMSa without concern, up to obtain the rest of elements of the Groebner basis.
 \item[iii)] When we have finished the algorithm, for each $b\in \L$ or $(b,c)\in \L\times \L$ we obtain the corresponding polynomials  $e_b$ or $e_{(b,c)}$. Then we check if $\H$ is afforded for it. 
 \item[iv)] It is clear that one and only one of all possible $e_b$ or $e_{(b,c)}$ would be the desired generator polynomial.
\end{itemize}

Finally, one may see that we may extend this procedure for more than one missing value.

\end{document}